\documentclass{emulateapj}
\usepackage{amsmath,natbib}
\usepackage{graphics,graphicx}

\newcommand\chandra{{\sl Chandra}}
\newcommand\xmm{{\sl XMM}}

\newcommand\xspec{{\sc xspec}}

\newcommand\ergcms{{\rm erg cm$^{-2}$ s$^{-1}$}}

\begin{document} 

%\title{WHIM in the Sculptor Wall Detected via X-Ray Absorption}
%\author{David A. Buote, Taotao Fang, \& Philip J. Humphrey}  
%\affil{Department of Physics and Astronomy, University of
%California at Irvine, 4129 Frederick Reines Hall, Irvine, CA 92697-4575}
\title{X-Ray Absorption By WHIM in the Sculptor Wall}
\author{David A. Buote\altaffilmark{1}, Luca Zappacosta\altaffilmark{1,2},
Taotao Fang\altaffilmark{1}, Philip
J. Humphrey\altaffilmark{1}, \\ Fabio Gastaldello\altaffilmark{1,3,4},
\& Gianpiero Tagliaferri\altaffilmark{5}}
\altaffiltext{1}{Department of Physics and Astronomy, University of
California at Irvine, 4129 Frederick Reines Hall, Irvine, CA 92697-4575}
\altaffiltext{2}{INAF - Osservatorio Astronomico di Trieste, Via Tiepolo 11,
34143 Trieste, Italy}
\altaffiltext{3}{INAF - IASF Milano, Via E. Bassini 15, I-20133 Milano, Italy}
\altaffiltext{4}{Occhialini Fellow}
\altaffiltext{5}{INAF-Osservatorio Astronomico di Brera, Via Bianchi,
46, 23807 Merate, Italy}

\slugcomment{Accepted for Publication in The Astrophysical Journal}

\begin{abstract}
We present \xmm\ RGS and \chandra\ LETG observations of the blazar,
H~2356-309, located behind the Sculptor Wall, a large-scale galaxy
structure expected to harbor high-density Warm-Hot Intergalactic
Medium (WHIM).  Our simultaneous analysis of the RGS and LETG spectra
yields a $3\sigma$ detection of the crucial redshifted \ion{O}{7}
K$\alpha$ line with a column density ($\ga 10^{16}$~cm$^{-2}$)
consistent with similar large-scale structures produced in
cosmological simulations. This represents the first detection of
non-local WHIM from X-ray absorption studies where \xmm\ and \chandra\
data are analyzed simultaneously and the absorber redshift is already
known, thus providing robust evidence for the expected repository of
the ``missing baryons''.
\end{abstract}


\section{Introduction}
\label{intro}

At most 50\% of the total baryonic matter in the nearby universe can
be accounted for by the amount of luminous baryons revealed by
observations of stellar light, narrow Ly$\alpha$ absorption systems,
and X-ray emission from hot gas in galaxy clusters
\citep[e.g.,][]{fuku98}. Cosmological simulations predict that most of
the ``missing baryons'' (30-50\%) reside in low-density plasma in
large-scale filamentary structures between galaxies
\citep[e.g.,][]{cen99,dave01}. The location and predicted temperature
range ($10^5-10^7$~K) of this plasma motivated its name as the
``Warm-Hot Intergalactic Medium'' (WHIM).

Ultraviolet observations of \ion{O}{6} absorption lines in background
quasar spectra provided the first clear detection of the WHIM
\citep[e.g.,][and references therein]{sava98,trip06}. The \ion{O}{6}
lines probe lower temperature WHIM ($\approx 10^5$~K), while most of
the WHIM gas is expected to exist at higher temperatures observable
only at X-ray wavelengths \citep[e.g.,][]{dave01}. Important evidence
for WHIM in X-ray emission has been provided by, e.g., correlation
studies of X-ray emission and galaxy positions in superclusters
\citep[e.g.,][]{zapp02,zapp05} and by X-ray imaging studies of the
highest density WHIM near massive clusters \citep[e.g.,][and
references therein]{wern08}.

Because of the difficulty in treating the foreground X-ray emission
from the Milky Way, it is important to corroborate these detections
with X-ray absorption-line studies which are largely insensitive to
this issue. Typically, X-ray absorption studies focus on the
\ion{O}{7} K$\alpha$ line ($\lambda = 21.6$~\AA), because it is
expected to be the strongest line at X-ray energies owing to the
anticipated WHIM temperature and the relatively high cosmic abundance
of oxygen.

To date there is no detection of WHIM in X-ray absorption that is
generally accepted by the astronomical community \citep[e.g.,][and
references therein]{rich08}. The case of Mkn~421 illustrates the
problem. \citet{nica05} first reported highly significant detections
of two absorption systems along the Mkn~421 sight line using a deep
\chandra\ observation. But
\citet{kaas06} showed that the statistical significance of the lines
reported by \citet{nica05} was greatly overestimated because the
redshifts of the absorbers were not known {\it a priori}. Furthermore,
a deeper \xmm\ observation of Mkn~421 did not confirm the presence of
the claimed absorption systems \citep{rasm07}.

To address these issues, we consider a different observational
approach proposed by \citet{krav02} and pioneered by \citet{fuji04} in
their absorption study of the Virgo cluster. Rather than performing a
so-called ``blind'' search for previously unknown intervening WHIM
structures in front of the brightest background AGN, we perform a
``targeted'' search by choosing a region of sky that provides an
optimal combination of known large-scale structure and a bright,
intrinsically featureless, background source. Because the redshift of
the foreground structure is known {\it a priori}, the statistical
significance of a detected line is much higher than for a blind
search. By focusing on large-scale structure, we should observe
relatively high density WHIM, therefore aiding a detection. We
restrict our study to background blazars because of their
intrinsically featureless spectra compared to other AGN.

\begin{figure}[t]
\begin{center}
\includegraphics[scale=0.5,angle=0]{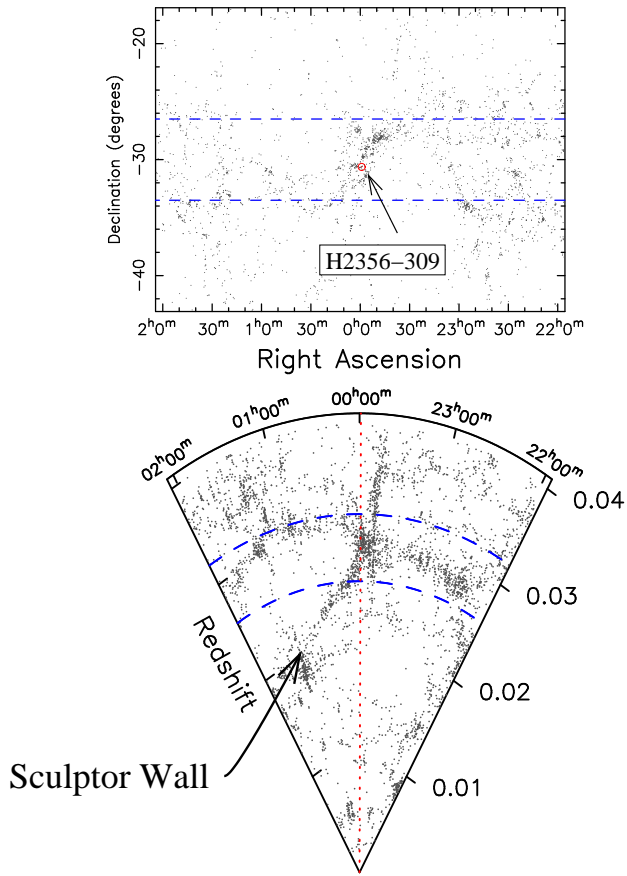}
\end{center}
%\vskip -0.6cm
\caption{\footnotesize 
Sky map (top) and wedge diagram (bottom), each expressed in R.A., of
the region of the Sculptor Wall where the blazar H~2356-309 is
located. Galaxy redshifts are taken from NED. The (blue) dashed lines
in the upper panel define the range of declination of the galaxies
displayed in the lower panel.  Conversely, the (blue) dashed lines in
the lower panel define the redshift range of the points in the upper
panel. The blazar line-of-sight is indicated by the (red) dotted line
in the wedge diagram.}
\label{sky}
%\vskip -0.6cm
\end{figure}

In this paper we present \xmm\ and \chandra\ grating observations of
the blazar H~2356-309 ($z=0.165$) located behind the Sculptor Wall
(Figure \ref{sky}), a southern super-structure filled with many groups
and clusters of galaxies \citep{daco94}. The part of the Sculptor Wall
intercepted by the sight-line of H~2356-309 represents a redshift
range of 0.028-0.032. Although the observations were designed as
Targets-of-Opportunity to catch H~2356-309 in outburst, the apparent
outbursts were not long enough to enable the follow-up (10-40 hours
after triggering) to catch the blazar in a high state.  The
observations were performed four months apart because of different
triggering criteria used for the different satellites.

Despite H~2356-309 being observed in its low state ($\sim
10^{-11}$~erg~cm$^{-2}$~s$^{-1}$ 0.5-2.0~keV flux), we detected a
candidate WHIM \ion{O}{7} resonance line in the Sculptor Wall from a
simultaneous fit of the \xmm\ and \chandra\ data. Although we examined
other candidate lines from other parts of the spectra using all the
\xmm\ and \chandra\ data, here we present only the results for the
\ion{O}{7} line because only for it do we achieve a detection of at
least $3\sigma$ significance.

The paper is organized as follows. In \S \ref{obs} we present the
observations and describe the data preparation. The spectral fitting,
modeling approach, and systematic errors are discussed in \S
\ref{spec}. We present our conclusions in \S \ref{conc}.

\section{Observations}
\label{obs}

On June 2, 2007 \xmm\ observed H~2356-309 (ObsID 0504370701) for
approximately 130~ks with the Reflection Grating Spectrometer (RGS) as
the primary instrument. The RGS is comprised of two nominally
identical sets of gratings, the RGS1 and RGS2. But due to the failure
of one of the CCD chips early in the mission, the RGS2 lacks
sensitivity over the energy range ($20-24\, \AA$) relevant for the
\ion{O}{7} lines. Consequently, we focus on the RGS1 in this paper.
We analyzed the RGS1 data with the most up-to-date XMM-Newton Science
Analysis Software (SAS
8.0.0)\footnote{http://xmm.esac.esa.int/sas/8.0.0/} along with the
latest calibration files.

We generated a light curve of the background events using CCD number
9; i.e., the CCD that is close to the optical axis and is most likely
to be affected by the background flares.  Inspection of the light
curve does reveal a flare near the end of the observation, which we
removed, resulting in a clean exposure of 126~ks for the
RGS1. Using these good time intervals, we reprocessed the RGS1 data to
produce the data files required for spectral analysis; i.e., the
response matrix, the background spectrum, and the file containing the
spectrum of H~2356-309 (which also contains background).  We restrict
our analysis to the first-order spectra, because the second order does
not cover the relevant \ion{O}{7} line energies and its count rate is
much lower.

\chandra\ observed H~2356-309 for 95.5~ks with the Low-Energy
Transmission Grating (LETG) and the HRC-S detectors, which offer the
best compromise between sensitivity and spectral resolution in the
wavelength range of the \ion{O}{7} lines. We reduced the data using
the standard \chandra\ Interactive Analysis of Observations (CIAO)
software (v4.0) and \chandra\ Calibration Database (CALDB, v3.4.0) and
followed the standard \chandra\ data reduction
threads\footnote{http://cxc.harvard.edu/ciao/threads/}. To ensure
up-to-date calibration, we reprocessed the data from the ``level 1''
events file to create a new ``level 2'' file for analysis.  We applied
the lastest HRC gain map and pulse-height filter for use with LETG
data\footnote{http://cxc.harvard.edu/contrib/letg/GainFilter/}. This procedure
removed a sizable portion of background events with negligible X-ray
event loss. From inspection of the light curves extracted from
source-free regions of the detector, we concluded that the observation
was not affected significantly by background flares.

We used the CIAO software to produce the files required for spectral
analysis; i.e., the response matrix, the background spectrum, and the
file containing the spectrum of H~2356-309 (which also contains
background). Unlike the RGS, different orders cannot be separated
from the LETG-HRC-S spectra, and so a given spectrum contains all
orders. Consequently, by default we use a response matrix that
includes information up to the sixth order. In \S \ref{sys} we compare
results using a response matrix that includes information only on the
first-order spectrum.

We re-binned the spectra to optimize detection of the absorption
lines. After some experimentation, we achieved this by requiring a
minimum of 75 and 40 counts per bin respectively in the RGS1 and LETG
spectra.

\begin{figure*}[t]
\begin{center}
\includegraphics[scale=0.45,angle=-90]{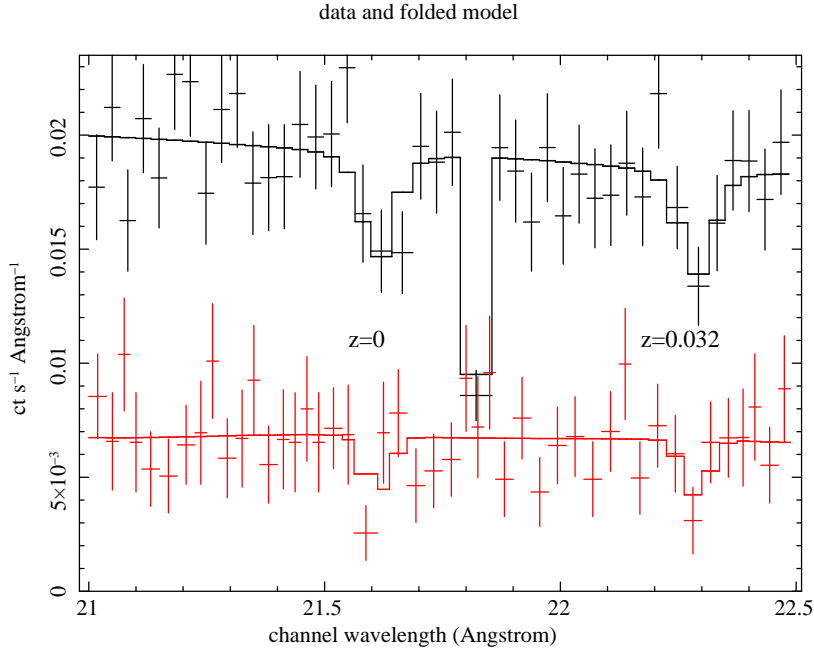}
\end{center}
%\vskip -0.6cm
\caption{\footnotesize  
Background-subtracted \xmm\ (black) and \chandra\ (red) spectra of the
blazar H~2356-309 over the wavelength range $21.0-22.5$\AA\ used to
study the candidate \ion{O}{7} lines. The spectral models represent a
power-law continuum, two absorption lines broadened by the
instrumental resolution and the Voigt function, and foreground
Galactic absorption from cold gas. The wavelength positions of the two
absorption lines, nearby and Sculptor, are indicated.}
\label{spectrum}
%\vskip -0.6cm
\end{figure*}

In Figure \ref{spectrum} we present the background-subtracted
\chandra\ and \xmm\ spectra. (The background subtraction is achieved in the normal way in
xspec via the "data" and "backgrnd" commands.) Two candidate
absorption lines in each detector are immediately apparent upon visual
inspection. The line near 21.6\AA\ corresponds to \ion{O}{7} for
$z=0$, consistent with nearby WHIM in the Local Group or hot gas in
the Milky Way. However, the other candidate line near 22.3\AA\
corresponds to \ion{O}{7} for $z\approx 0.03$, consistent with the
redshift of the unvirialized structure in the Sculptor Wall. (The
absorption line near 21.8\AA\ in the RGS1 is a known instrumental
feature.)

\section{Spectral Fitting}
\label{spec}

We performed spectral fitting using \xspec\ v11.3.2ag
\citep{xspec} and minimized the Cash C-statistic \citep{cstat}, rather
than $\chi^2$, to obtain unbiased parameter estimates from the
Poisson-distributed spectral data \citep{hump08c}. In all the fits the
foreground absorbing Hydrogen column density from cold gas in the
Milky Way was fixed at the value \citep[$1.33\times
10^{20}$~cm$^{-2}$;][]{dick90} using the {\sc phabs} model in \xspec.

\subsection{Models}
\label{mod}

A common practice in X-ray studies of the WHIM is to characterize the
properties of a candidate absorption line by subtracting a gaussian
component from a model of the continuum. The equivalent width is
computed from this composite model, from which the column density of
the absorber is inferred. This procedure is well suited to the case
where the absorber is optically thin (i.e., linear part of the curve
of growth), which we show is not the case for the candidate Galactic
and extra-galactic WHIM absorbers in H~2356-309. In these cases a
composite continuum-minus-gaussian model takes negative values at the
line centers. Consequently, we constructed a physical line absorption
model as follows.

Let $I(E;0)$ be the monochromatic specific intensity of photon energy
$E$ incident at coordinate $s=0$ assuming plane-parallel geometry. The
emergent radiation at coordinate $s$ is then,
\begin{equation}
I(E;s) = I(E;0)e^{-\tau(E;s)},
\end{equation}
where the optical depth\footnote{We do not correct for stimulated
emission because (1) the effect is small for the resonance
\ion{O}{7} line for expected WHIM temperatures, (2) the WHIM
temperature is not precisely determined by our fits, and (3) the
$(1-\exp[-E/K_BT])$ correction term applies assuming a
thermal plasma, which may not be strictly valid for the WHIM.} is,
\begin{equation}
\tau(E;s) = {\rm N}(s) \frac{\pi e^2}{m_e c}f \phi(E).
\end{equation}
The quantity, ${\rm N}(s) =
\int_0^s n{\rm ds}$, is the column density of absorbing atoms, and $f$
is the absorption oscillator strength of the spectral line
transition. The line-profile function,
\begin{equation}
\phi(E) = \frac{h}{\Delta E_D \sqrt{\pi}} {\rm H}(a,u),
\end{equation}
is expressed in terms of the Doppler width,
\begin{equation}
\Delta E_D = E_0\frac{b}{c},
\end{equation}
where $E_0$ is the rest-frame energy of the line, and
\begin{equation}
b = \sqrt{\frac{2k_BT}{m}} = 32.24\sqrt{\frac{T}{10^6\rm K}
\frac{16}{A}} \, \, \, \frac{\rm km}{\rm s},
\end{equation}
is the Doppler $b$-parameter. We have expressed $b$  in
terms of an oxygen atom of 16 atomic mass units. The quantity ${\rm H}(a,u)$
is the Voigt function which combines the effects of natural and
Doppler broadening through the parameters,
\begin{equation}
a = \frac{h \Gamma}{4\pi\Delta E_D}, \hskip 1cm  u = \frac{\Delta E}{\Delta E_D},
\end{equation}
where, $\Gamma$ is the Einstein-A coefficient of the line transition,
and $\Delta E = E - E_0$. We used the implementation of the Voigt
function provided by \citet{wells99}
\footnote{http://www.atm.ox.ac.uk/user/wells/voigt.html}.  We
implemented the line absorption as a local multiplicative model in
\xspec, such that for a given energy bin of an input response matrix we evaluated
the average of $\exp[-\tau(E;s)]$ over 10 equally spaced energies
within the bin. (The variation of $I(E;0)$ over an energy bin in the
RGS and LETG response matrices is negligible.)  For the resonance
\ion{O}{7} K$\alpha$ line we used $E_0=0.574$~keV (21.6\AA),
$f=0.696$, and $\Gamma = 3.3\times 10^{12}$~s$^{-1}$ \citep{vern96b}.

We find that a power-law model is a good description of the continuum
over $21.0-22.5$~\AA\ ($0.5515-0.5909$~keV) in both \chandra\ and
\xmm. Both the photon spectral index and the normalization are fitted
separately for each detector. Two absorption lines, one near $z=0$ and
another near $z=0.03$, were added following the above prescription. We
required the parameters of the absorption lines to be the same in both
detectors. Therefore, the model that we fitted to the data consists of
a power-law continuum, two absorption lines, and (fixed) foreground
Galactic absorption from cold gas.

\subsection{Continuum}
\label{cont}

Upon fitting the composite model simultaneously to the \xmm\ and
\chandra\ spectra we obtain a spectral index of $3.3\pm 1.7$
and a 21.0-22.5~\AA\ flux of $(7.76 \pm 0.26)\times 10^{-13}$~\ergcms
(unabsorbed) for the power-law continuum of the \xmm\ spectrum (90\%
errors quoted for both quantities). Because the \chandra\ LETG--HRC-S
spectrum contains photons from all orders, not just the first order,
some photons with wavelengths smaller than 21.0~\AA\ in the
higher-order spectrum will be mis-classified as having wavelengths
between $21.0-22.5$~\AA\ if the response matrix only includes
information on the first-order diffraction spectrum. To account for
the higher orders, the response matrix (\S \ref{obs}) requires the
spectral model to be defined down to wavelengths as small as $\sim
5$~\AA.

We adopted the following procedure to define the continuum for the
\chandra\ data as a compromise between the desire to insure an
accurate modeling of photons from higher orders and to measure the
continuum using only data very close to the energies of the absorption
lines. We initially fitted the power-law model modified by foreground
Galactic absorption from cold gas to the background-subtracted
\chandra\ data over the broad wavelength range $5-30$~\AA. This model
provides a very good fit to the \chandra\ spectrum with a
well-constrained power-law index, $1.883\pm 0.034$. In all subsequent
fits (though see \S \ref{sys}) to the narrow band (21.0-22.5~\AA) we
restrict the the power-law index for the \chandra\ data to lie between
these values established by the broad-band fit, thereby insuring an
accurate modeling of the broad-band spectrum and thus the contribution
from the higher orders. After fitting the composite model (i.e.,
continuum, absorption lines, foreground cold absorber) simultaneously
to the \xmm\ and \chandra\ spectra we obtain a 21.0-22.5~\AA\ flux of
$5.52^{+0.39}_{-0.46}\times 10^{-13}$~\ergcms (unabsorbed) for the
\chandra\ power-law continuum; i.e, the continuum flux of the blazar
measured by \xmm\ is about 40\% higher than observed by \chandra\ over
the spectral region of study.

\subsection{Background}
\label{bkg}

Over $21.0-22.5$~\AA\ the background comprises about 8\% of the
observed \xmm\ count rate and about 38\% of the \chandra\ count
rate. As noted above, we perform spectral fitting on the
background-subtracted \xmm\ and \chandra\ spectra of the
blazar. However, below in \S \ref{lines} we also assess the
statistical significance of the \ion{O}{7} absorption lines using
simulated spectra, for which we prefer to use model representations of
both the source and background spectra for each satellite. We
constructed models of the background spectra as follows.

The \chandra\ LETG-HRC-S background spectrum over $5-30$~\AA\ is
fairly well described by a broken power-law folded through only the
instrument RMF (``redistribution matrix file'') and not the ARF
(``auxiliary response file''). This is achieved by defining a ``{\sc
bknpow}/b'' model in \xspec\ v11. The model parameters are very
tightly constrained. For reference, the best-fitting values for the
power-law indexes are, 1.552 and 1.888, and the best-fitting break
energy is, 1.338~keV (9.27~\AA).  Because the model is not a perfect
fit to the data at all wavelengths, we adjusted the flux of the
power-law to best match the spectral region of the oxygen lines. That
is, adopting the values of the spectral indexes and break energy
obtained from the broad-band fit we refitted the background model over
the restricted range $21.0-22.5$~\AA\ of interest, which yields a
normalization that is $4\% \pm 2\%\, (1\sigma)$ less than obtained for
the broad-band fit.

For the \xmm\ RGS1 the broad-band spectrum is not well described by
simple models. Consequently, we adopted a single power-law model,
again folded only through the RMF (``{\sc pow}/b'' model in \xspec\
v11), but over a narrower wavelength region where the model is a good
fit. We selected the wavelength range, $20-24$~\AA, as the largest
range producing a good fit while enclosing the critical
$21.0-22.5$~\AA\ interval. We obtain a power-law spectral index of
1.16 with a 90\% confidence range (0.27-3.65).

\subsection{Absorption Lines}
\label{lines}

\begin{table}[t] \scriptsize
\begin{center}
\caption{O~{\rm VII} Spectral Line Properties}
\label{data}
\begin{tabular}{lccc}  \tableline\tableline\\[-7pt]
& & $\rm N$\\
& Redshift & $(\rm 10^{16}\, cm^{-2})$ & $\tau_0$\\
\tableline \\[-7pt]
Milky Way & -0.0004 - 0.0020 & $(5.5,0.9)$ & $(20.3,2.2)$ \\
Sculptor & 0.0311 - 0.0327 & $(4.7,1.0)$ & $(18.7,2.1)$\\
\tableline \\
\end{tabular}
\tablecomments{All parameters are quoted at the 90\%
confidence level. The redshift ranges are determined by allowing the
Doppler $b$-parameter to take any value between 1-300 km/s, a large
range that should bracket physical WHIM values, and the redshifts
themselves were allowed to take any values during the fits.  The
column densities and line-center optical depths are lower limits
evaluated assuming two values for the $b$-parameter $(50,100)$ km/s.}
\end{center}
\end{table}

In this section we discuss the significance and properties of the
absorption lines obtained from fitting the composite model to the
data, while systematic errors in our analysis are examined in \S
\ref{sys}.

We list in Table \ref{data} the redshifts of the two candidate
\ion{O}{7} K$\alpha$ absorption lines obtained from fitting the
spectral model simultaneously to the \xmm\ and \chandra\ data. The
line near $21.6$~\AA\ is fully consistent with hot gas from the Milky
Way (or Local Group WHIM), with the inferred error range on the
redshift being consistent with zero. The redshift range of the line
near $22.3$~\AA\ matches very well the range of structure in the
Sculptor Wall ($z=0.028-0.032$) that intercepts the blazar's
line-of-sight. Henceforth, we refer to these candidate lines as the
Galactic and Sculptor WHIM lines.

The change in the fit statistic, whether it be the C-statistic or
$\chi^2$, is very similar when adding the Galactic and Sculptor WHIM
lines, indicating that the statistical significance of the lines is
also similar. For reference, the change in the $\chi^2$ statistic
observed when adding the Sculptor WHIM line to the fit suggests the
line is significant at the $\approx 3\sigma$ level when interpreted in
terms of the F-Test. However, the F-Test is not strictly applicable to
spectral line detection \citep{prot02}.

To obtain a rigorous estimate of the statistical significance of the
Sculptor WHIM line we employed a Monte Carlo procedure. Using as a
reference the best-fitting spectral model excluding the Sculptor WHIM
line, we generated with \xspec\ synthetic \xmm\ and \chandra\ spectra
for the source and background (\S \ref{bkg}) with the same exposure
times as the real data. The simulations were performed using the full
energy resolution of the response matrices, after which the synthetic
spectra were re-grouped as described in \S
\ref{obs}. In this way the simulations account for noise associated
with the re-binning procedure.

We fitted the synthetic background-subtracted spectra with the
reference model and minimized the C-statistic to obtain a best
fit. Then we added a second absorption line component {\it with
redshift range restricted to 0.028-0.032} appropriate for the Sculptor
Wall. After obtaining the new best fit, we recorded the reduction in
the C-statistic. We performed this procedure for $10^4$ simulations.
The probability of a false detection is given by the number of
simulations where the reduction in the C-statistic is at least as
large as observed in the real fit relative to the total number of
simulations.

The fit of the reference model (i.e., without the Sculptor WHIM line)
to the real data yields a best-fitting value of the C-statistic of
68.2 for a total of 88 spectral bins. Upon adding the Sculptor line
and re-fitting, we obtain a C-statistic of 58.6 for a reduction of
9.6. For the Monte Carlo procedure described above, we find that in
only 36 out of the $10^4$ simulations is the C-statistic reduced by at
least 9.6. This translates to a false detection probability of
0.36\%. Therefore, the statistical significance of the Sculptor WHIM
line is 99.64\%; i.e., about $3\sigma$. Following the same procedure,
but this time starting without the Milky Way line and with the
Sculptor WHIM line already in place, we determine the statistical
significance of the Milky Way line to be 99.69\%, also about $3\sigma$.

(We mention that the \xmm\ data contribute more to the statistical
significance than do the \chandra\ data. If the spectral fitting is
performed separately for each data set, then following the above
Monte Carlo procedure for the Sculptor WHIM line we obtain a statistical
significance of $2.4\sigma$ for \xmm\ and $1.7\sigma$ for \chandra.)

Apart from the redshift noted above, the line parameters are not well
constrained by the data. In Table \ref{data} we list the 90\%
confidence lower limits obtained for the \ion{O}{7} K$\alpha$ column
density (N) and line-center optical depth $(\tau_0 = \tau(E_0,s))$. We
quote results for two values of the $b$-parameter
$(50,100)$~km~s$^{-1}$ which span a large range of WHIM temperature
and turbulent pressure.

The lower limits of $\tau_0\approx 1$ for both lines demonstrate that
the line centers are not optically thin. Therefore, one cannot infer
the column density from the equivalent width independent of the
$b$-parameter; i.e., these lines do not lie on the linear portion of
the curve of growth. For reference, the best-fitting equivalent widths
are nearly 30~m\AA\ for both the Galactic and Sculptor lines.

The column density lower limits of $\approx 10^{16}$~cm$^{-2}$ are
consistent with expectations for both lines. Similar column densities
previously have been reported for the Milky Way (or Local Group WHIM)
\citep[e.g.,][]{fang03}. As for the Sculptor line, we re-examined the
cosmological hydrodynamical simulation of \citet{cen06} and identified
five super-structures with similar (or higher) galaxy over-density
relative to the Sculptor Wall. By shooting random sight-lines through
these super-structures we obtain a mean \ion{O}{7} column density of
$\approx 3\times 10^{16}$~cm$^{-2}$, consistent with the limits
imposed by our spectral fits on the Sculptor Wall.

\subsection{Systematic Errors}
\label{sys}

The large statistical errors obtained for the line column densities
and line-center optical depths dominate systematic errors associated
with our data analysis and modeling choices, and therefore we do not
provide a detailed systematic error budget for the line
parameters. However, in this section we briefly discuss the possible
impact of systematic errors on the estimation of the significance of
the Sculptor Wall WHIM line, which is the principal result of our
paper. We have in particular considered the effects of errors in the
background models and the wavelength calibration of the detectors, and
we have also examined using $\chi^2$ rather than the C-statistic.

As we mentioned in \S \ref{bkg}, the background level is only a small
fraction of the total observed flux for the \xmm\ RGS. Consequently,
it is expected that small variations in the background level from that
we adopted in our fits should have little effect on the significance
estimate. Indeed, varying the background level by $\pm 5\%$, which
reflects the statistical error on the power-law model obtained for the
RGS background, did not have a noticeable effect.

Because the \chandra\ LETG-HRC-S is unable to separate spectra of
different orders, we examined how much our results change if instead
we used a response matrix containing only information on the
first-order spectrum. Using only the first-order response leads to an
underestimation of the actual background by almost 20\% over the
$21.0-22.5$~\AA\ range. Since the first-order response does not require
a valid continuum model outside our range of interest, we allowed the
slope of the continuum to be fitted without restriction over
$21.0-22.5$~\AA, which provides a measure of the sensitivity of the
results to the definition of the continuum. When performing
simultaneous RGS-LETG fits using the first-order response for the LETG
we obtain a statistical significance of the Sculptor WHIM line of
$99.53\%$, only slightly less than obtained when using the more
accurate response containing orders 1 through 6.

We also considered systematic uncertainties in the wavelength
calibration of the detectors. Current estimates indicate the
calibrations are accurate to within 2.4~m\AA\ in the
RGS1\footnote{A. Pollock, XMM-Newton Users Group Meeting
Presentations, 6-7 May 2008, http://xmm.esac.esa.int/external/
xmm\_user\_support/usersgroup/20080506/} and to 10~m\AA\ in the
LETG/HRC-S\footnote{M.\ Chung et al.,
http://cxc.harvard.edu/cal/Letg/Corrlam/}. We find little difference
in any of our results when allowing for shifts of these magnitudes
between the detectors.  We do mention that if the wavelength of the
$z=0$ line is allowed to be fitted separately for the RGS1 and the
LETG then the best-fitting line centers are offset by
48~m\AA. However, the shift is significant only at the $\sim 90\%$
confidence level, and it is substantially larger than the expected
calibration uncertainty, indicating this marginal shift is probably
noise. Therefore, we kept the line energies tied between the
detectors.

Finally, we also performed all of our fits by minimizing
(data-weighted) $\chi^2$ and obtained a consistent result for the
significance of the Sculptor line as obtained using the
C-statistic. In particular, the statistical significance of the
Sculptor line obtained using $\chi^2$ is, 99.81\%, about $3\sigma$.

\section{Conclusions}
\label{conc}

We report a detection at the $3\sigma$ level of an \ion{O}{7}
K$\alpha$ resonance absorption line located in the Sculptor Wall
super-structure of galaxies.  The column density ($\ga
10^{16}$~cm$^{-2}$) of the line, though not precisely constrained, is
consistent with that produced by similar structures generated in
cosmological simulations. The presence of this absorption line is
inferred from simultaneous analysis of \xmm\ RGS1 and \chandra\
LETG/HRC-S spectra of the blazar H~2356-309 $(z=0.165)$.  Since the
line center is saturated (for both the Sculptor and Local lines), we
interpreted the spectra with a physical model where the line is
modified by the Voigt profile to account for both natural and Doppler
broadening.

The robustness of the detection benefits from three key aspects of our
study. First, a critical factor in obtaining this high detection
significance level is that we know {\it a priori} the redshift of the
intervening structure in the Sculptor Wall ($z=0.028-0.032$). The
importance of this point was dramatized in the controversy over the
claimed WHIM detection in the spectrum of Mrk~421
\citep{nica05,kaas06}. Second, the line is detected from a joint
analysis of \xmm\ and
\chandra\ data. The consistent picture from the two satellites
provides important reassurance given again the controversy for Mrk~421
where the RGS did not confirm the claimed \chandra\ detection
\citep{rasm07}.  Third, since the background source is a
blazar, it is very unlikely that features intrinsic to the source
contaminate our study of the foreground WHIM.

This detection of WHIM in the Sculptor Wall represents the most
significant (non-local) WHIM detection to date from X-ray absorption
studies, providing vital evidence for the expected repository of the
``missing baryons'' \citep[e.g.,][]{fuku98,cen99,dave01} complementary
to that of X-ray emission studies of WHIM in superclusters
\citep[e.g.,][]{zapp05} and the intersections of binary clusters
\citep[e.g.,][]{wern08}, and \ion{O}{6} absorption line studies that
probe WHIM at lower temperatures \citep[e.g.,][]{sava98,trip06}.

\acknowledgements 

We thank the referee for re-analyzing the \xmm\ data and verifying our
result.  D.A.B., T.F, and P.J.H.\ gratefully acknowledge partial
support from NASA through Chandra Award Numbers GO7-8140X issued by
the Chandra X-ray Observatory Center, which is operated by the
Smithsonian Astrophysical Observatory for and on behalf of NASA under
contract NAS8-03060. We also are grateful for partial support from
NASA-XMM grant NNX07AT24G. We thank Dr.\ R.\ Remillard for providing
daily RXTE fluxes for H~2356-309. This research has made use of the
NASA/IPAC Extragalactic Database (NED) which is operated by the Jet
Propulsion Laboratory, California Institute of Technology, under
contract with the National Aeronautics and Space Administration.

%XXX bibtex bibliography \\
%\bibliographystyle{apj}
%\bibliographystyle{apj_hyper}
%\bibliography{dabrefs}

\end{document}